\documentclass{moriond}
\usepackage{amsmath}
\usepackage{bm}

\def\be{\begin{equation}}
\def\ee{\end{equation}}

\newcommand{\Photo}{\includegraphics[height=35mm]{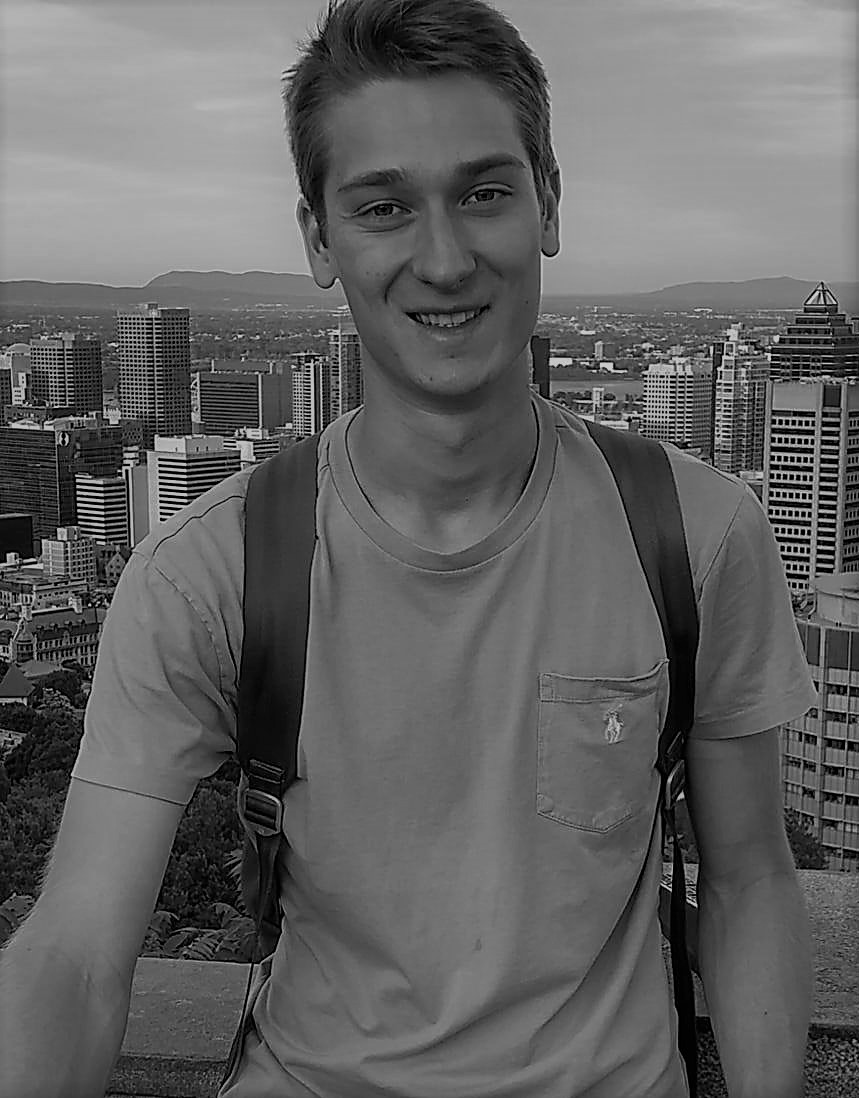}}

\newcommand{\beq}{\begin{equation}}
\newcommand{\eeq}{\end{equation}}
\newcommand{\bal}{\begin{aligned}}
\newcommand{\eal}{\end{aligned}}
\newcommand{\bea}{\begin{eqnarray}}
\newcommand{\eea}{\end{eqnarray}}
\def\be{\begin{equation}}
\def\ee{\end{equation}}

\def\beq{\begin{equation}}
\def\eeq{\end{equation}}

\newcommand{\K}{\mathcal K}

\renewcommand{\L}{\mathcal L}

\def\be{\begin{equation}}
\def\ee{\end{equation}}
\def\ba{\begin{eqnarray}}
\def\ea{\end{eqnarray}}

\def\d{\mathrm{d}}

\usepackage[normalem]{ulem}

\def\ba{\begin{eqnarray}}
\def\ea{\end{eqnarray}}

\def\L{\mathcal{L}}
\def\K{\mathcal{K}}
\def\X{\mathcal{X}}

\newcommand{\Ostro}{{Ostrogradski }}

\def\d{\mathrm{d}}
\def\mn{_{\mu \nu}}
\def\mnup{^{\mu \nu}}

\def\mupn{^\mu_{\phantom{\mu}\nu}}
\def\({\left(}
\def\){\right)}

\def\p{\partial}

\begin{document}
\vspace*{4cm}
\title{COSMOLOGY OF A NEW CLASS OF MASSIVE VECTOR FIELDS}

\author{ V. POZSGAY }

\address{Theoretical Physics, Blackett Laboratory, Imperial College,\\ London, SW7 2AZ, UK}

\maketitle\abstracts{
I introduce (Extended) Proca-Nuevo, a non-linear theory of a massive spin-1 field enjoying a non-linearly realized constraint. I will provide a covariantization scheme that allows for consistent, ghost-free cosmological solutions, describing the correct number of dynamical variables in the presence of perfect fluid matter. I will finally exhibit explicit hot Big Bang solutions featuring a late-time self-accelerating epoch, where all the stability and subluminality conditions are satisfied and where gravitational waves behave precisely as in General Relativity.}

\section{Introduction}

Finding the most general theory of a massive vector field can play an important part in the ongoing search for extensions to General Relativity (GR). It was proven in the context of Generalized Proca (GP) theories \cite{DeFelice:2016yws,DeFelice:2016uil} that a massive vector condensate could behave as Dark Energy (DE). Here we provide a new and non-equivalent massive vector field theory, dubbed as Proca-Nuevo (PN) \cite{deRham:2020yet} (and its extension, namely Extended Proca-Nuevo (EPN) \cite{deRham:2021efp}), that provides an interesting DE model and predicts a self-accelerating branch for a hot Big Bang scenario. 


\section{Proca-Nuevo and Extended Proca-Nuevo}
\label{sec:ExtPN}

We start with a vector field $A_\mu$ living on flat spacetime with Minkowski metric $\eta\mn$. The construction of PN theory follows the intuition drawn from the helicity decomposition of massive gravity \cite{deRham:2011qq}, beginning with the definition
\ba
	f\mn[A] = \eta\mn + 2 \frac{\p_{(\mu} A_{\nu)}}{\Lambda^2} + \frac{\p_{\mu} A^\rho \p_{\nu} A_\rho}{\Lambda^4} \,,
		\label{eq:deff2}
\ea
where $\Lambda$ is an energy scale that will ultimately control the strength of the vector self-interactions. Next we introduce the tensor $\K\mupn$ defined as \cite{deRham:2010kj,deRham:2014zqa}
\ba
\label{eq:defK1}
 \K\mupn &=& \X\mupn -\delta \mupn \,, \\
 \text{with }	\quad \X\mupn[A] &=& \left( \sqrt{\eta^{-1}f[A]}  \right)\mupn \,, \qquad \text{i.e.}\qquad \X^{\mu}_{\phantom{\mu} \alpha} \X^{\alpha}_{\phantom{\alpha} \nu} = \eta^{\mu \alpha}f_{\alpha \nu} \,.
		\label{eq:defK2}
\ea
In four dimensions, the PN and EPN theories for the vector field $A_\mu$ are then expressed as \cite{deRham:2020yet,deRham:2021efp}
\beq
	\L_{\text{PN}} = \Lambda^4 \sum_{n=0}^4 \alpha_n(X) \L_n[\K[A]] \,,\qquad \L_{\text{EPN}} = \Lambda^4 \sum_{n=0}^4 \alpha_n(X) \L_n[\K[A]] + \Lambda^4 \sum_{n=1}^4 d_n(X) \frac{\L_n[\p A]}{\Lambda^{2n}} \,,
	\label{eq:LextEPNflat}
\eeq
where the $n$th order symmetric polynomial of a matrix $M$ is defined by
\begin{equation}
	\L_n[M] = - \frac{1}{(4-n)!} \epsilon^{\mu_1 \cdots \mu_n \mu_{n+1} \cdots \mu_4} \epsilon_{\nu_1 \cdots \nu_n \mu_{n+1} \cdots \mu_n} M^{\nu_1}_{\phantom{\nu_1}\mu_1} \cdots M^{\nu_n}_{\phantom{\nu_n}\mu_n}\,.
\label{eq:defLnK}
\end{equation}
In Eq. \ref{eq:LextEPNflat} the coefficients $\alpha_n(X)$ and $d_n(X)$ are arbitrary functions of
\beq
X=-\frac{1}{2\Lambda^2}\,A^{\mu}A_{\mu}\,.
\eeq
Note that the product $\alpha_0(X)\L_0\equiv V(A^{\mu}A_{\mu})$ contains the standard potential of the vector field. In order for the trivial vacuum $\langle A_{\mu}\rangle=0$ to be a consistent state one should demand that $\alpha_0$ have a non-zero quadratic contribution, i.e.\ $\alpha_0\supseteq -\frac 12 (m^2/\Lambda^4) A^{\mu}A_{\mu}$.

\paragraph{Null Eigenvector.}

In PN (and EPN) theory the constraint is realized through a field-dependent null eigenvector (NEV). Indeed, it was shown that there exists a non-perturbative normalized time-like NEV $V_{\mu}^{\rm PN}$ of the PN operators satisfying $\eta^{\mu \nu}V^{\rm PN}_{\mu} V^{\rm PN}_{\nu} = -1$ and $\mathcal{H}^{\mu \nu} V^{\rm PN}_{\mu} = 0$. The NEV $V_{\mu}^{\rm PN}$ and the Hessian matrix of time derivatives $\mathcal{H}^{\mu\nu}$ are given by \cite{deRham:2020yet}
\begin{equation} \label{eq:PN NEV}
	V_{\mu}^{\rm PN}(\Lambda) = (\mathcal{X}^{-1})^{0 \nu} \(\eta\mn+\frac{1}{\Lambda^2}\p_\nu A_\mu\) \,, \qquad \mathcal{H}^{\mu \nu}=\frac{\partial^2\L_{\rm PN}}{\partial\dot{A}_{\mu}\partial\dot{A}_{\nu}}\,.
\end{equation}
Given that the additional  $d_n(X) \L_n[\p A]$ operators do not affect the Hessian matrix, the NEV of EPN coincides with that of PN defined above in Eq. \ref{eq:PN NEV}.


\section{Special model without non-minimal couplings}
\label{sec:SpecEx}

In the present work we focus on a covariantization in which all non-minimal coupling terms are omitted. The field $A_{\mu}$ lives on a spacetime with metric $g\mn$, $\nabla$ corresponds to covariant derivatives with respect to that metric, and $f_{\mu\nu}$ and $\K\mupn$ are promoted to covariant operators by taking $\p \rightarrow \nabla$ and $\eta \rightarrow g$. The ``special'' model we consider is defined by the action
\begin{equation}
	S = \int \d^4 x \sqrt{-g} \left(\frac{M_{\rm Pl}^2}{2}\,R + \L_{\rm EPN}^{\rm (cov)} + \L_M \right) \,.
	\label{eq:Shat}
\end{equation}
Here, $R$ is the curvature scalar, and the covariant EPN Lagrangian reads
\begin{equation}
\L_{\rm EPN}^{\rm (cov)} = - \frac14 \,F\mnup F\mn + \Lambda^4 \left( \L_0^{\rm (cov)} + \L_2^{\rm (cov)} + \L_2^{\rm (cov)} + \L_3^{\rm (cov)} \right) \,,
\end{equation}
where
\begin{equation}
\begin{alignedat}{2}
\L_0^{\rm (cov)} &= \alpha_{0}(X) \,,\qquad &&\L_1^{\rm (cov)} = \alpha_{1}(X) \L_1[\K] + d_{1}(X) \frac{\L_1[\nabla A]}{\Lambda^2} \,, \\
\L_2^{\rm (cov)} &= \alpha_{2,X}(X) \left(\L_2[\K] - \frac{\L_2[\nabla A]}{\Lambda^4}\right) \,,\qquad &&\L_3^{\rm (cov)} = - \frac16 \alpha_{3,X}(X) \left(\L_3[\K] - \frac{\L_3[\nabla A]}{\Lambda^6} \right) \,.
\end{alignedat}
\end{equation}
Finally, we choose $\L_M$ to be the Schutz-Sorkin perfect fluid matter Lagrangian \cite{SCHUTZ19771}, given by
\begin{equation}
	\L_M = - \rho_M(n) - \frac{J^{\mu}}{\sqrt{-g}} \left( \p_{\mu} l + \mathcal{A}_i \p_{\mu} \mathcal{B}^i \right) \,,\qquad \text{with} \qquad n= \sqrt{\frac{J^{\mu} J_{\mu}}{g}}  \,.
	\label{eq:SMSS}
\end{equation}

Note that the model has no non-minimal couplings between the vector field and the metric. This special model is free of ghost on cosmological backgrounds and this holds here upon accounting for the dynamical mixing between the gravitational and vector degrees of freedom.

\subsection{Background}
\label{ssec:Background}

We proceed  by deriving the background cosmological equations of motion.
We focus on the FLRW metric
\begin{equation}
	\d s^2 = -N^2(t)\d t^2 + a^2(t) \delta_{ij} \d x^i \d x^j \,,
	\label{eq:lineFLRW2}
\end{equation}
with the vector field profile
\begin{equation} \label{eq:vector bkgd2}
	A_{\mu}\d x^{\mu} = -\phi(t)\d t \,.
\end{equation}

\begin{figure}[!htb]
	\center{\includegraphics[height=5cm]{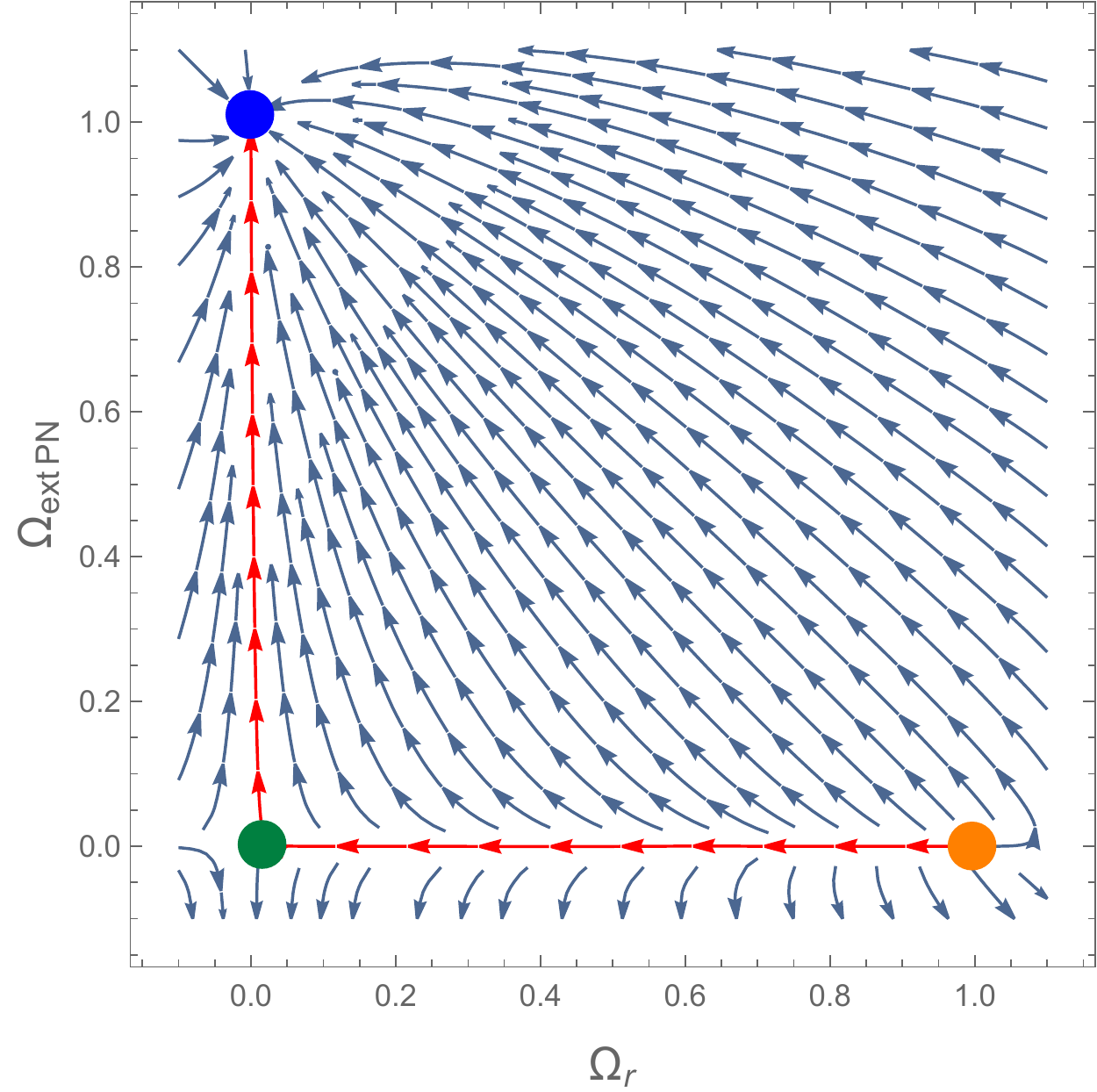}~~~~~~~~\includegraphics[height=5cm]{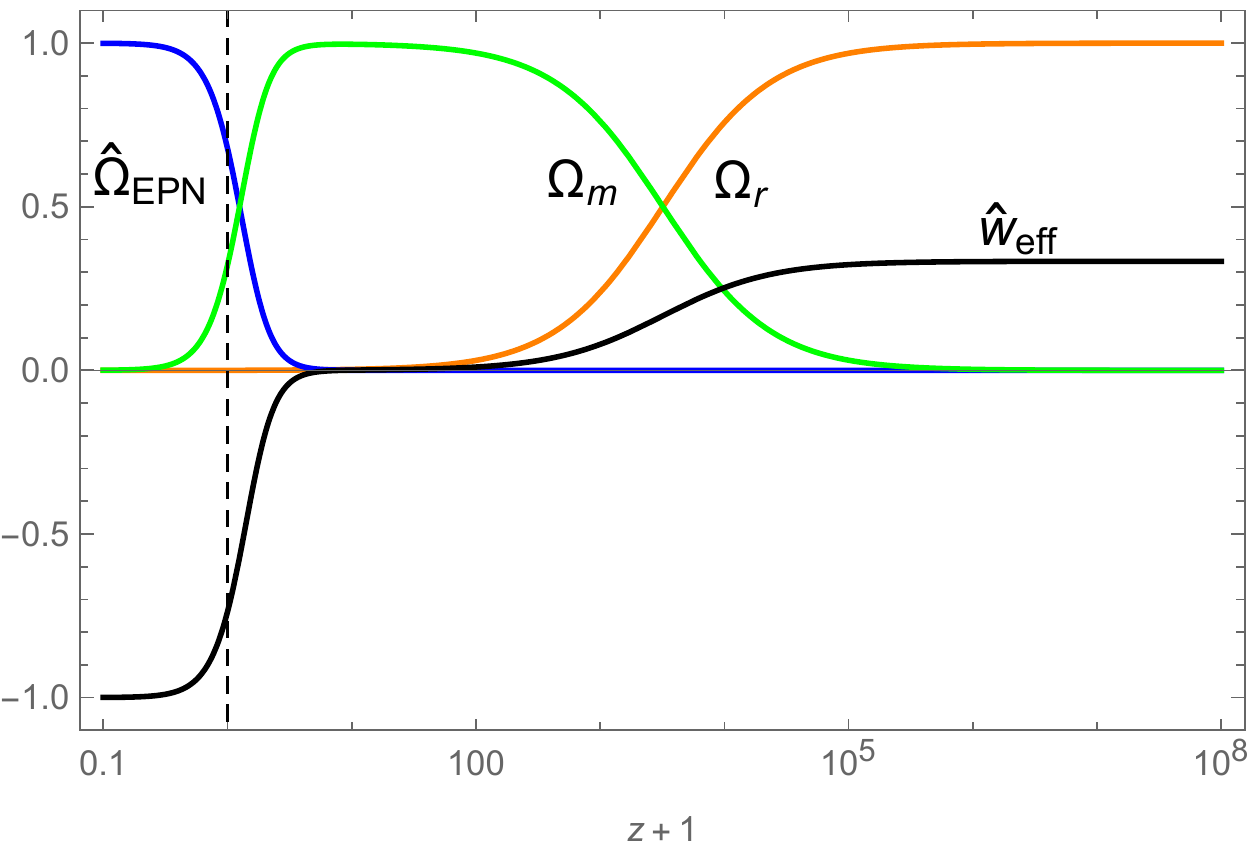}}
	\caption{\label{fig:plot_phase_portrait} (left) Phase portrait of the autonomous system Eq. \ref{eq:dynsys}. The red trajectory is a particular solution resembling the hot Big Bang phase of our universe with epochs of radiation (orange dot), matter (green) and dark energy (blue) domination. (right) Evolution of the density parameters as functions of redshift, with initial conditions chosen such that $\Omega_{\text{EPN}} = 0.68$ and $\Omega_r = 10^{-4}$ at $z=0$. Taken from \protect\cite{deRham:2021efp}.}
	\label{fig:fig1}
\end{figure}

We specify the matter perfect fluid to be a mixture of pressureless matter ($P_m=0$) and radiation ($P_r = \frac13 \rho_r$), respectively denoted by subscripts ``$m$'' and ``$r$'', i.e.\ $\rho_M = \rho_m + \rho_r$ and $P_M = P_m + P_r$. This means that the equations of state translate into $\dot{\rho}_m + 3H \rho_m = 0$ and $\dot{\rho}_r + 4H \rho_r = 0$. Next, it is convenient to introduce the density parameters
\begin{equation}
	\Omega_r \equiv \frac{\rho_r}{3 M_{\text{Pl}}^2 H^2}\,, \qquad \Omega_m \equiv \frac{\rho_m}{3 M_{\text{Pl}}^2 H^2}\,, \qquad \Omega_{\text{EPN}} \equiv \frac{\rho_{\text{EPN}}}{3 M_{\text{Pl}}^2 H^2} \,,
\end{equation}
where $\rho_{\text{EPN}}$ is the effective energy density of the vector condensate, so that the Friedmann equation reads
\begin{equation}
	\Omega_r + \Omega_m + \Omega_{\text{EPN}} = 1 \,.
\end{equation}
We now specify the parameters of the model by making the following choice:
\ba \label{eq:special model choice of coeffs}
&	\alpha_0 = - \frac{m^2}{\Lambda^2} X \,, \qquad
	\alpha_1 = - \frac{\Lambda^4}{M_{\text{Pl}}^4} b_1 X^2 - \frac{\Lambda^2}{M_{\text{Pl}}^2} c_1 X \,,\qquad   d_1 =- \frac{\Lambda^4}{M_{\text{Pl}}^4} e_1 X^2 + \frac{\Lambda^2}{M_{\text{Pl}}^2} c_1 X \,, \\
&	\alpha_{2,X} = \frac{\Lambda^4}{M_{\text{Pl}}^4} b_2 X^2 + \frac{\Lambda^2}{M_{\text{Pl}}^2} c_2 X \,,\quad {\rm and}\quad
	\alpha_{3,X} = \frac{\Lambda^4}{M_{\text{Pl}}^4} b_3 X^2 + \frac{\Lambda^2}{M_{\text{Pl}}^2} c_3 X \,, \\
&	b_1 + e_1 = \frac{1}{12\sqrt{6}} \,,\qquad m^2 M_{\text{Pl}}^2 = \Lambda^4 \,.
\ea
We are interested in the time evolution of the density parameters $\Omega_{\text{EPN}}$ and $\Omega_r$ ($\Omega_m$ being trivially determined from these). Primes denote derivatives with respect to the e-folding number $N=\log(a)$. The autonomous system determining the evolution of $\Omega_{\text{EPN}}$ and $\Omega_r$ reads
\beq\bal
		\Omega_{\text{EPN}}' &= \frac{4 \Omega_{\text{EPN}} ( 3(1-\Omega_{\text{EPN}}) + \Omega_r) }{3 + \Omega_{\text{EPN}}} \,,\\
		\Omega_r' &= - \frac{\Omega_r \left( 3 ( 1 - \Omega_r ) + 13 \Omega_{\text{EPN}} \right)}{3 + \Omega_{\text{EPN}}} \,.
	\label{eq:dynsys}
\eal\eeq
A straightforward analysis shows that this system admits three fixed points corresponding to radiation domination, matter domination and dark energy domination (de Sitter (dS) fixed point). There exists a particular solution that qualitatively mimics our universe's hot Big Bang phase, starting very close to the radiation point, flowing toward the matter point, and then asymptotically approaching the dS point, as seen in Fig. \ref{fig:fig1}.

\subsection{Perturbations}
\label{ssec:Perturbations}

One can now add perturbations to the metric, the vector field $A_{\mu}$ and the matter content. The perturbations include tensor, vector and scalar sectors, decoupling from each other at the quadratic level in the action. It is possible to prove that the tensor sector is identical to the one of GR. This means that EPN predicts the same gravitational waves as GR. 

Turning then to the vector and scalar sector, it is also possible to show that there is one helicity-$1$ and one helicity-$0$ mode that physically propagate. This amounts to the expected $3$ degrees of freedom of a healthy non-ghostly massive vector field. Note that there is also a physical scalar and vector mode in the matter sector. 

It is possible to show that for some given choice of coefficients $b_I$, $c_I$ and $e_1$, one can get stable and subluminal perturbations. Explicitly, the kinetic terms are positive (avoiding \Ostro-instabilities), the square velocities are positive (avoiding gradient stabilities) and subluminal, and the mass terms are positive (avoiding tachyonic instabilities).


\section{Conclusions}
\label{sec:Conclusions}

We have proven that EPN is ghost-free on a flat background. We have then provided a consistent covariantization scheme for cosmological backgrounds. The background equations of motion can be solved to find that EPN dynamically coupled to gravity predicts a universe with three distinct domination era : radiation, matter, and finally dark energy. This is qualitatively in agreement with the observations. Furthermore, EPN provides a hot big-bang scenario with a self-accelerating branch, i.e. the acceleration is driven by the vector condensate and not a cosmological constant. Finally, all perturbations can be made stable and subluminal, giving a well-behaved and ghost-free theory on cosmological backgrounds, even at the level of perturbations. In particular, we predict gravitational waves to be exactly luminal, as in GR, which is in agreement with tight experimental bounds placed on their speed of propagation.

\section*{Acknowledgments}

V.P. would like to thank Claudia de Rham, Sebastian Garcia-Saenz and Lavinia Heisenberg.

\section*{References}

\bibliography{references}

\end{document}